\documentclass[aps,pra,twocolumn,superscriptaddress,10pt]{revtex4-2}
\usepackage{mathtools}
\usepackage{graphicx}
\usepackage{epstopdf}
\usepackage{lmodern}
\usepackage{xfrac}
\usepackage{natbib}
\usepackage{hyperref}
\usepackage{verbatim}
\usepackage{mathtools}

\usepackage{float}

\begin{document}
\title{Distributing Polarization Entangled Photon Pairs with High Rate over
  Long Distance through Standard Telecommunication Fiber
}

\author{Lijiong~Shen}
\affiliation{Centre for Quantum Technologies, National University of Singapore, 3 Science Drive 2, Singapore 117543}

\author{Chang~Hoong~Chow}
\affiliation{Centre for Quantum Technologies, National University of Singapore, 3 Science Drive 2, Singapore 117543}

\author{Justin~Yu~Xiang~Peh}
\affiliation{Centre for Quantum Technologies, National University of Singapore, 3 Science Drive 2, Singapore 117543}

\author{Xi~Jie~Yeo}
\affiliation{Centre for Quantum Technologies, National University of Singapore, 3 Science Drive 2, Singapore 117543}

\author{Peng~Kian~Tan}
\affiliation{Centre for Quantum Technologies, National University of Singapore, 3 Science Drive 2, Singapore 117543}

\author{Christian~Kurtsiefer}
\affiliation{Centre for Quantum Technologies, National University of Singapore, 3 Science Drive 2, Singapore 117543}
\affiliation{Department of Physics, National University of Singapore, 2 Science Drive 3, Singapore 117551}

\email[]{christian.kurtsiefer@gmail.com}
\date{\today}

\maketitle
\textbf{Entanglement distribution over long distances is essential for many
  quantum communication schemes like quantum teleportation, some variants of
  quantum key
  distribution~\cite{Bennett1993,BB84,ekert:91,Bennett1992a}, or
  implementations of a quantum internet~\cite{kimble2008quantum,pompili2021realization}. Distributing
  entanglement through standard telecommunication fiber is particularly
  important for quantum key distribution protocols with low vulnerability over
  metropolitan distances~\cite{Lutkenhaus2000,Acin2007,Scarani2009}. However,
  entanglement distribution over long distance through optical fiber
  so far could only be accomplished with  moderate photon pair
  rates~\cite{Vanner2007,Treiber2009,Wengerowsky2019,Wengerowsky2020a}. In
  this work, we present entanglement distribution over 50\,km of standard
  telecommunication fiber with pair rate more than 10,000\,s$^{-1}$ using a
  bright non-degenerate photon pair source. Signal and idler wavelengths of
  this source are optimized for low dispersion in optical fiber and high efficiency for  single-photon avalanche diode detectors, respectively.  
The resulting modest hardware requirement and high rate of detected entangled
photon pairs could significantly enhance practical entanglement-based quantum key distribution in existing metropolitan fiber networks.}

In recent years, the rapid development of quantum technologies has driven
Quantum Key distribution (QKD)
towards higher key generation rates~\cite{Zhang2009} and longer distances ---
metropolitan area~\cite{Tang2016,shi2020stable}, inter-city~\cite{Muller1996}, and even
inter-continental quantum communication has been demonstrated
recently~\cite{Liao2017}. Currently, building secure quantum networks on top
of these point-to-point quantum communication schemes is of increasing
interest in research and commercial applications~\cite{Chen2021}.

Most implementations of QKD are based on prepare-and-measure protocols due to
its optical simplicity. In contrast, entanglement-based QKD can offer the
advantage of not requiring a trusted active random number
generator~\cite{Marcikic:2006}, and is
immune to photon number splitting attacks, exposing fewer vulnerabilities in
a practical implementation~\cite{Lutkenhaus2000}. Moreover, entanglement is
also the basis of device-independent quantum cryptography, a class of
higher-security QKD schemes that  does not rely on the premise that the
involved devices can 
be trusted~\cite{Acin2007,Scarani2009}. Thus, it is highly desirable to
distribute entanglement over long distance for future quantum secure
networks.

While distributing entanglement over very long distance through free space~\cite{Yin2012} or even satellite links~\cite{Yin2017} is naturally appealing due to the low optical transmission loss in air, 
optical fibers can guide photons over the long distance without challenging
alignments or environmental considerations. They also are the sole option when
a line-of-sight is unavailable. Optical fibers are a particularly attractive
transmission method in metropolitan areas, where ITU G.652D standard compliant fiber is likely already deployed as part of their existing telecommunication infrastructure.

Early work has demonstrated  distribution of  polarization-entangled photons
through fiber over 1.45\,km~\cite{Poppe2004} for relatively short wavelengths.
Polarization entanglement is much easier to prepare, measure, and couple to
other physical systems compared to time-bin
entanglement~\cite{Tittel1998}, which is a 
common encoding scheme at telecom wavelengths because it requires fewer
expensive detectors. Time-bin entanglement is also less susceptible to
polarization degradation due to polarization mode dispersion (PMD) in optical
fibers, an effect that is severe for broadband photons found in many
sources of polarization-entangled photons~\cite{Ribordy2000}. 
Improved technology continuously lowered the PMD of optical fibers, and
permitted to manufacture low loss optical filters to reduce photon bandwidth. Consequently, polarization entanglement distribution  over tens or even more than 100\,km optical fiber has been demonstrated in recent years~\cite{Vanner2007,Treiber2009,Wengerowsky2019,Wengerowsky2020a}.

However, these demonstrations had to make optimizations that limit their
practicality.
For example, dispersion-shifted
fibers (typically G.655 standard compliant fibers) have lower
transmission losses and dispersion at 1550\,nm compared to G.652D
fibers~\cite{C.Inc.2019}, but are far less deployed in existing
telecommunication networks. Alternatively, chromatic dispersion and PMD could be
reduced by using narrow band pass filters together with conventional entangled
pair sources, but this reduces the heralding efficiency of photon pairs even
for ideal spectral filters with unity transmission in pass
band~\cite{PhysRevA.95.061803}, eventually leading to a lower entangled
photon pair detection rate. Consequently, most reported work on long distance
entanglement distribution was carried out with photon pair rates that are too
low for most practical applications.
Similarly, superconducting nanowire single-photon detectors (SNSPDs) are used
in many demonstrations of long distance entanglement distribution due to their
high efficiency, low dark counts, and low timing
jitter~\cite{Natarajan2012}. However, SNSPDs need to be cooled below 4\,K,
which requires a significant overhead in power, size, and cost in comparison
with typical single-photon avalanche photodiodes (APDs).

In this work, we present a highly non-degenerate entangled photon pair source
based on type-0 spontaneous parametric down-conversion (SPDC) process to
overcome these drawbacks in fiber-based entanglement distribution.
One of the photons in pairs has a wavelength of 1310\,nm and falls in the
zero-dispersion window of G.652D standard compliant fiber, and a
sub-nanometer bandwidth even without any spectral filtering. The other
photon in the pairs has a wavelength around 586\,nm where Si-APDs have
nearly optimal detection efficiency. With such a source, we observe an average
entanglement visibility of more than 97$\%$ locally. After transmitting the
1310\,nm  photons through 50\,km  of G.652D compliant fiber, we still observe
a photon pair coincidence rate over 10,000 s$^{-1}$  with a pair of APDs, with
a raw polarization entanglement visibility of 92.5$\%$. The corresponding
estimated quantum key rate in a BBM-92 like protocol~\cite{Bennett1992a} with such a source would
be 5,172 s$^{-1}$. This is a significant boost in rates for practical quantum key distribution in a metropolitan area.

\begin{figure}[t]
\includegraphics[width=1\linewidth]{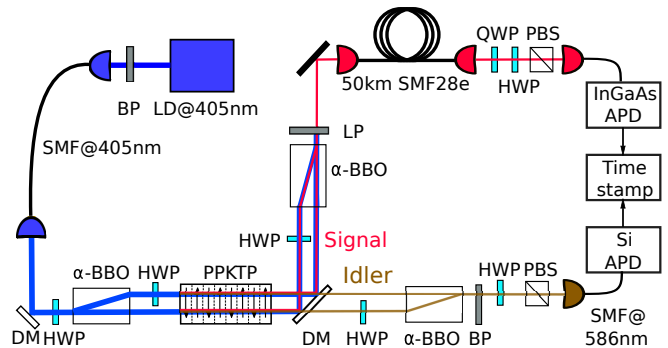}
\caption{\label{fig:setup} 
Schematic of the non-degenerate entangled photon pair source. Pump light at
405\,nm is split into two paths and is subject to type-0 SPDC inside the PPKTP
crystal. The polarization state in the upper path is rotated by 90 degrees by
a half-wave plate. Each pump path generates photon pairs with a non-degenerate
wavelength  at 1310\,nm (signal), and 586\,nm (idler). A dichroic mirror (DM)
separates the non-degenerate photons from both pump paths. A pair of
$\alpha$-BBO crystals recombine the two signal and two idler paths separately to create an entangled state $|\Phi^-\rangle$=$\frac{1}{\sqrt{2}}(|HH\rangle-|VV\rangle)$. 
A motorized HWP and a PBS projects idler photons into different
polarizations. Signal photons at 1310\,nm are first coupled into a long
SMF28e+ telecommunication fiber before projection into linear bases in a
similar way as the idler photons. 
}
\end{figure}

A sketch of the experimental setup is shown in Fig.~\ref{fig:setup}.
Time-correlated photon pairs are generated via SPDC in a periodically poled
potassium titanyl phosphate crystal (PPKTP, 2 $\times$ 1 $\times$ 20\,mm$^3$)
that is pumped with light from a grating-stabilized diode laser at a
wavelength of 405\,nm.
Polarization-entangled photon pairs are prepared into a $|\Phi^-\rangle$
state by allowing two
indistinguishable conversion processes separated spatially in the PPKTP
crystal~\cite{Fiorentino:08,lohrmann2020broadband}. Three $\alpha$-BBO
crystals act as beam splitter/displacers to interfere these conversion
processes. 
An InGaAs-APD and a Si-APD
detect signal and idler photons of the entangled pairs within a
narrow coincidence window. Details can be found in Methods.

\begin{figure}[t]
\includegraphics[width=1.15\linewidth]{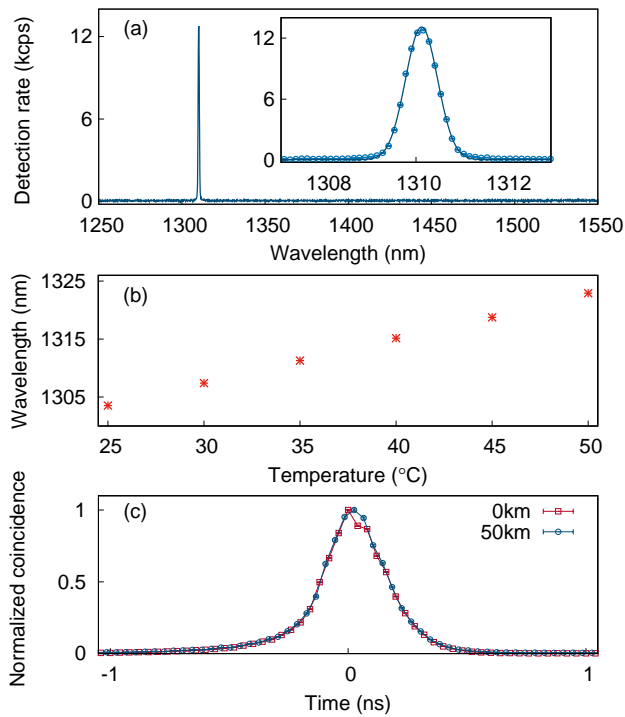}
\caption{\label{fig:spec} 
(a) The spectrum of signal photons from the non-degenerate pair source at a
temperature of 33.4\,$^\circ$C of the  PPKTP crystal shows a single peak
around 1310\,nm with a nearly Gaussian profile (see fit to data in Inset). (b)
Signal photon wavelength dependence on PPKTP temperature. (c) Time correlation
of photon detection events from a pair of APDs, with the signal photon passing
through 0\,km and 50\,km of SMF28e+ fiber, respectively.  The temporal
correlations for the 50\,km transmission case are offset and normalized for
visual comparison. Both histograms show a nearly indistinguishable
distribution with a FWHM around 280\,ps. }
\end{figure}

\begin{figure}[t]
\includegraphics[width=1\linewidth]{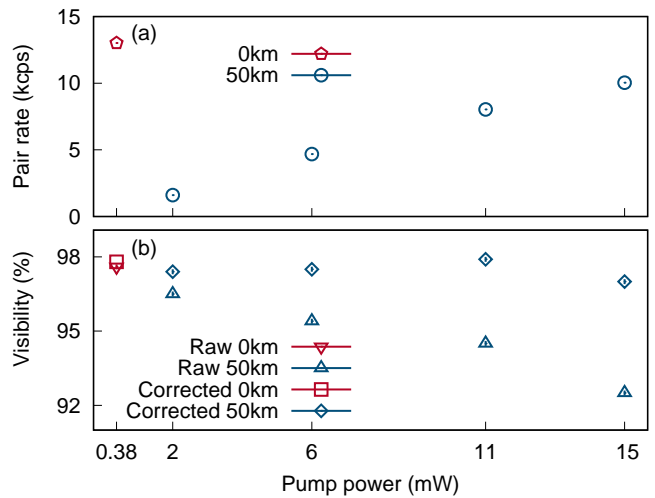}
\caption{\label{fig:power} 
(a) Average pair rates for different pump powers. The locally collected photon
pair rate, averaged over H/V and D/A bases, is 13,012$\pm$14\,s$^{-1}$ at
380\,$\mu$W pump power. Pair rates with the 50\,km fiber in the signal arm are
$\text{1,606}\pm3$\,s$^{-1}$, 4,686$\pm$6\,s$^{-1}$, 8,033$\pm$10\,s$^{-1}$, and
10,033$\pm$16\,s$^{-1}$ at their corresponding pump powers.
(b) Visibility of polarization correlations as a measure of entanglement
quality for different pump powers.
The raw visibility observed locally with 380\,$\mu$W pump is 97.6$\pm$0.1$\%$.
After propagating through 50\,km of SMF28e+ fiber, the raw visibility reduces to
96.5$\pm$0.1$\%$,  95.4$\pm$0.1$\%$, 94.5$\pm$0.1$\%$,  and 92.5$\pm$0.1$\%$
for a respective pump power of 2, 6, 11, and 15\,mW.  Background-corrected
visibilities range from 97\,$\%$ to 98\,$\%$ for all the measurements.
}
\end{figure}

The pair generation rate of the source is first characterized without the
entanglement mechanism mediated by the beam displacers.
At a pump power of 100\,$\mu$W, a pair rate of more than 10,000 s$^{-1}$ is
observed with over 20$\%$  heralding efficiency; this value increases to 22$\%$ after correction for detector dark counts and after-pulses. It is worth  noting that even with a relatively optimistic estimation of the detector efficiency
(60$\%$ for the Si-APD, and 15$\%$ for the InGaAs-APD), the ideal heralding
efficiency assuming a perfect fiber coupling while ignoring losses on multiple
optical elements is predicted to be $\sqrt{(15\%\times60\%)}=30\%$ (see Methods). This implies a close to optimal
collection efficiency.
The single rate at the Si-APD exceeds 70,000 s$^{-1}$ with a pump power of
100\,$\mu$W. Assuming  a Si-APD efficiency of 60$\%$, we can safely infer that
at a pump power of 1\,mW, the source should generate more than 1 million
photon pairs in a second; a direct measurement would have severely saturated the
available detectors.

Figure~\ref{fig:spec}(a) shows a spectrum of the signal photons recorded with a
grating spectrometer, with only a 
single peak in a range of 300\,nm, with a center wavelength of 1310.12\,nm and
a Full Width at Half Maximum (FWHM) of 0.84\,nm obtained from a fit to a
Gaussian distribution. The spectrometer resolution is separately determined to
be 0.47\,nm (FWHM) with a distributed feedback laser at 1310\,nm
and a bandwidth of 3\,MHz.
Thus, the de-convoluted bandwidth of the source signal photons at 1310\,nm is
around 0.7\,nm, corresponding to 120\,GHz.  Figure~\ref{fig:spec}(b) shows the
change of the center wavelength of signal photons with the temperature of the
PPKTP crystal. The wavelength increases almost linearly with a slope of
0.8\,nm/K over the zero-dispersion range (1304\,nm-1324\,nm) of the SMF28e+, G652D-compliant fiber.   

To experimentally verify the zero-dispersion property of the fiber on signal
photons, we carried out a coincidence measurement on photon pairs from the
source with and without the 50\,km fiber. For
better timing resolution, the Si-APD was replaced with a device with smaller
timing jitter ($<$40\,ps) but lower detector efficiency (40$\%$).  The
coincidence peak is shifted by around 247\,$\mu$s for the measurement with the
50\,km fiber. Both coincidence measurements are normalized to their maximum coincidence rate for easier visual comparison. Figure~\ref{fig:spec}(c) clearly shows that the dispersion is unnoticeable even with the fast Si-APD.

The entanglement quality measurements were first carried out locally
with the three $\alpha$-BBO beam displacer inserted in the source, and
polarization projections in signal and idler paths are implemented as shown in
Fig.~\ref{fig:setup}. To characterize entanglement, we
extracted the visibility of polarization correlations in different bases,
details are explained in Methods.
The additional optics slightly reduces the source heralding efficiency from 22$\%$ to 18$\%$ for both pump paths.
 At 380\,$\mu$W pump power,  a raw visibility of   99.7$\pm$0.1$\%$ for the
 horizontal/vertical (H/V) basis, and 95.5$\pm$0.1$\%$ for the
 diagonal/anti-diagonal (D/A) basis is measured without 50\,km delay fiber, resulting in an average raw visibility of 97.6$\pm$0.1$\%$ (97.8$\pm$0.1$\%$ after correcting for background coincidences, see Methods). The corresponding coincidence rate averaged over H/V and D/A bases is 13,012$\pm$14 s$^{-1}$.

Then, the 50\,km SMF28e+ fiber (-17\,dB loss) is inserted between the source
and the polarization analyzer on the signal branch. The observed coincidence rates
and visibilities for different pump powers are shown in
Fig.~\ref{fig:power}. As expected, the coincidence rate increases mostly
linearly with pump power. We attribute the small nonlinearity at
15\,mW pump power to an onset of saturation of the Si-APD  (at a detection
rate of  $3.8\times10^6$\,s$^{-1}$; specified maximum count rate $3\times10^7$\,s$^{-1}$).
The raw average visibility of the polarization correlation decreases with the
pump power due to the quadratically increasing background coincidence (lower
panel). However, after correcting the visibility for background coincidences,
the visibility of polarization correlations and hence the quality of
entanglement remains at an almost constant high level with no strong dependence on the pump power.

\begin{figure}
\includegraphics[width=1\linewidth]{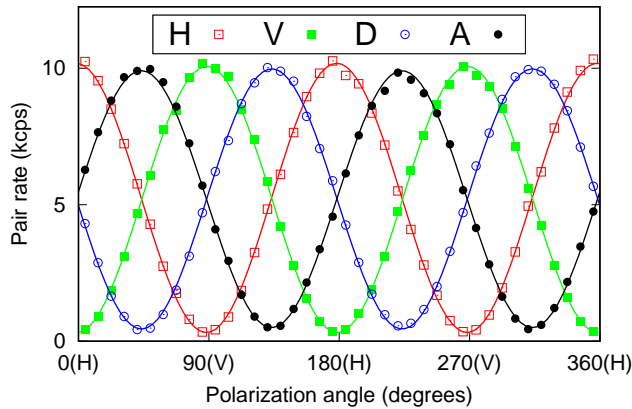}
\caption{\label{fig:visb} 
Polarization correlations in both H/V and D/A bases measured at 15\,mW pump
power after transmission of the signal photons through 50\,km SMF28e+
fiber. The coincidence rate is measured with polarization optics before signal and
idler photon detection, as shown in Fig.~\ref{fig:setup}. The polarization on
the idler side is kept at one of the four linear settings (H, V, D, A) while the
polarization on signal side is rotated over 360 degrees using the half
wave plate.}
\end{figure}

The raw visibility at 15\,mW pump power is  94.0$\pm$0.1$\%$ in the H/V basis,
and 91.0$\pm$0.2$\%$ in the D/A basis, corresponding to an average visibility
of 92.5$\pm$0.1$\%$ (see Fig.~\ref{fig:visb}). After correction for background
coincidences, the average visibility at 15\,mW pump power reaches
97.1$\pm$0.1$\%$. The  pair rate averaged over H/V and D/A bases is
10,033$\pm$16\,s$^{-1}$. Assuming that the measurement basis is chosen
randomly and uniformly, that noise parameters are independent of measurement
settings, and using a realistic bidirectional error correction efficiency
value of 1.1~\cite{Elkouss2011,Neumann2021}, the estimated key distribution
rate in a BBM-92 protocol would be 5,172 s$^{-1}$.
\newline

In conclusion, we  present an entangled photon-pair
source that is suitable for long distance quantum communication over the most
deployed optical fiber compliant with the G.652D standard.
The pair source generates signal photons at the zero-dispersion wavelength
(1310\,nm), while the corresponding idler photons have a wavelength of
586\,nm, which is close to the optimal efficiency of typical Si-APDs. 
The source has a brightness of more than $10^6$ pairs/s/mW. With the close to
ideal mode matching that can be accomplished with it, more than 10,000
s$^{-1}$ pairs are observed locally at 100\,$\mu$W pump power using a pair of
APDs. Entangled photon pairs can be generated by  beam-displacement
interferometer with minimal loss in brightness, and show a visibility of
polarization correlations of over 97\,$\%$.

The low intrinsic bandwidth of 0.7\,nm of signal photons in the conversion
process strongly suppresses temporal dispersion of photons passing through 50\,km SMF28e+, G652D-compliant optical
fiber, and permits to use a narrow coincidence time window which helps to
reject accidental coincidences that would deteriorate the observable entanglement quality. At the same time, an entangled photon pair rate of  more than  10,000
s$^{-1}$ with the same fiber can be detected with relatively simple avalanche photodetectors (APDs).
With the usual error correction methods, this source/detection combination can
be used to generate a secure bit rate of more than 5,000\,s$^{-1}$ in a BBM-92 protocol using simple
polarization encoding~\cite{Elkouss2011,Neumann2021}. With the presented source, entanglement distribution can  easily be extended to distances beyond 100\,km at a rate of hundreds of pairs/second, without switching to more sophisticated superconducting single-photon detectors. 
To our knowledge, the entangled pair rate is at least an order of
magnitude higher than other pair sources over the  similar distance, despite
the larger fiber attenuation per unit length in the O-band compared to the
C-band~\cite{Vanner2007,Treiber2009,Wengerowsky2019}.  We therefore believe
that such a source can significantly improve entanglement distribution over
metropolitan distances, permitting an integration of high-rate
entanglement-based quantum key distribution into existing telecommunication
network with realistic detection devices.

\bibliographystyle{naturemag}
\bibliography{ref}

\section*{Methods}
\textbf{Entanglement generation.}
Pump light is first generated by a grating-stabilized laser diode, and guided
through a single-mode fiber to clean up the spatial mode. A dichroic mirror then removes fluorescence
photons from the pump laser and fiber. An $\alpha$-BBO crystal splits the pump
light into two paths of orthogonal polarizations (horizontal, $H$, and vertical, $V$) with a beam separation of 1\,mm through birefringence.
A half-wave plate (HWP) transforms the $H$ polarized pump light part back to
$V$. The type-0 SPDC process in both conversion paths generates photon pairs in
a state $|VV\rangle$, which then are separated by wavelength using a second dichroic mirror. The polarization of the
separated signal and idler photons from one of the pump paths is transformed
to horizontal by two HWPs designed for their respective wavelengths. With
this, the crystal generates  $ |VV \rangle $ and $ |HH \rangle $ photon pairs
in the different pump paths of the same crystal, which are then spatially
overlapped by two $\alpha$-BBO crystal beam splitters, one in each collection
path to take care of the  
different birefringent properties at signal- and idler wavelengths.
The entanglement of photons arises from the indistinguishability between the photon pairs generated in the two down-conversion paths.
By adjusting the relative power and phase between the two pump paths, the maximally entangled state
\begin{align*}
|\Phi^-\rangle=\frac{1}{\sqrt{2}}(|HH\rangle - |VV \rangle)
\end{align*}
can be prepared.

\begin{figure}
\includegraphics[width=1\linewidth]{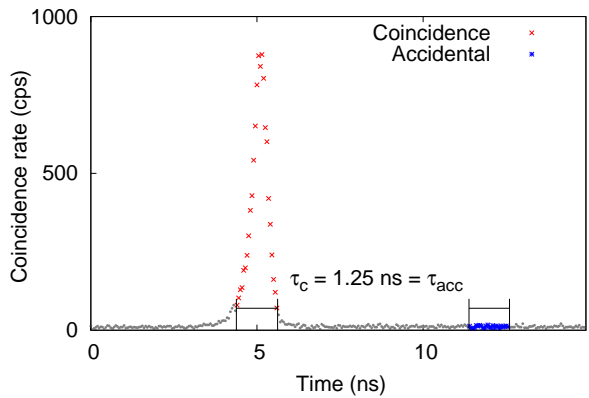}
\caption{\label{acc} 
Correlation of photon detection events from a pair
of APDs for entanglement distribution, with the signal photon passing through
50 km of SMF28e+ fiber. The horizontal axis for the time difference is offset
by about 247\,$\mu$s.  We consider detection events as coincidences when they
fall into the time interval $\tau_c$ (width: 1.25\,ns), and and measure the
coicnidence rate in a time window $\tau_{\text{acc}}$ of the same width,
displaced by about 7\,ns.
}
\end{figure}

Both a long pass (LP) filter with a cutoff wavelength at 1200\,nm
and a bandpass (BP) filter with a bandwidth of 20\,nm centered at 586\,nm
effectively remove pump light from the individual collection path. 
 A timestamp unit records the photon arrival time on both InGaAs-APD and Si-APD  to look for coincidences. The InGaAs-APD (IRSPD1 from S-Fifteen Instruments) and 
 the Si-APD (SPCM-AQRH-54 from Excelitas Technologies) have a timing jitter of  around 250\,ps and 500\,ps, respectively. We choose a coincidence window of 1.25\,ns 
 to capture the majority of the photon pairs for all the measurements reported
 in this work (see Fig.~\ref{acc}).

\textbf{Heralding efficiency.}
Assuming a photon pair production rate $P$ in the down conversion crystal,
single photon detectors with efficiencies
$\eta_{1}$ and $\eta_{2}$ will report single photon rates $S_{1(2)}=
P\eta_{1(2)}$, where $\eta_{1,2}$ are total
efficiencies, which include optical transmission, collection efficiencies
through mode matching, and detection efficiencies of the single-photon
detectors. A coincidence is registered when a photodetection event from each
detector fall in a coincidence window $\tau_c$.
 With negligible accidental coincidences, a coincidence rate $C$
 =$P$$\eta_{1}\eta_{2}$ can be observed.

The heralding efficiency $\eta$ of the photon pair is defined as the pair to
singles ratio~\cite{heff} $C$/$\sqrt{S_{1}S_{2}}$.
Substituting coincidence and single rates with the above expressions leads
to $\eta=\sqrt{\eta_{1}\eta_{2}}$.
Assuming ideal mode matching without any transmission loss, the  total efficiency $\eta_{1(2)}$ in each
collection path is just the corresponding detector efficiency.

\textbf{Visibility measurement.}
A combination of a HWP and a PBS (T$_\text{p}$:T$_\text{s}$ $>$ 1000:1) projects the polarization before coupling
into the spatial mode selection fiber for the idler photon (586\,nm). A
similar free-space polarization measurement scheme is used for signal photons,
where an additional quarter wave plate (QWP) is used to remove the ellipticity
after fiber passage; in this path, polarization elements HWP and PBS are also
placed behind the long fiber.
On the idler side, four different polarizations (horizontal, H, vertical, V,
diagonal, D, and anti-diagonal, A) are chosen by setting the HWP
accordingly.  On the signal side, the orientation of the linear polarization
is changed in finer steps by rotating the HWP to obtain a visibility of the
polarization correlations via $V=(C_{\text{max}}-C_{\text{min}})/(C_{\text{max}}+C_{\text{min}})$, where $C_{\text{max}}$ and $C_{\text{min}}$ are extracted from a sinusoidal fit to polarization correlations shown
in Fig.~\ref{fig:visb}.

{\bf Background correction.}
Apart from coincidences that can be attributed to photon pairs from the same
SPDC process, we detect accidental coincidences from different pair
generation processes, or from other detection processes. We estimate this
accidental coincidence rate $C_{\text{acc}}$ by recording the coincidence rate
for time differences displaced by 7\,ns from the main coincidence time
window (see Fig.~\ref{acc}). Then, the coincidence rates  are corrected by subtracting 
their respective accidental coincidence rate $C_{\text{acc}}$ at each HWP angle. The corrected visibility is evaluated as 
$V=(C_{\text{max}}^{\text{corr}}-C_{\text{min}}^{\text{corr}})/(C_{\text{max}}^{\text{corr}}+C_{\text{min}}^{\text{corr}})$. 
The $C_{\text{max}}^{\text{corr}}$ and $C_{\text{min}}^{\text{corr}}$ are also extracted from a sinusoidal fit to the corrected  polarization correlations.

\section*{}

\end{document}